\documentclass[cits]{PoS}

\newcommand{\BE}{\begin{equation}}
\def\EE{\end{equation}}
\def\BEA{\begin{eqnarray}}
\def\EEA{\end{eqnarray}}

\title{QCD equation of state at non-zero chemical potential }

\ShortTitle{QCD equation of state at non-zero chemical potential }

\author{S.~Basak,$^a$ 
A.~Bazavov, $^b$ C.~Bernard, $^c$ C.~DeTar,$^d$ W.~Freeman,$^b$
\speaker{Steven Gottlieb},$^a$
U.M.~Heller,$^e$
J.E.~Hetrick,$^f$ J.~Laiho,$^c$ L.~Levkova,$^d$ J.~Osborn,$^g$ R.~Sugar,$^h$ 
and D.~Toussaint$^b$\\
MILC Collaboration\\
        \llap{$a$} Department of Physics, Indiana University\\
	 Bloomington, IN 47405, USA\\
\llap{$^b$}
        Physics Department, University of Arizona\\ Tucson, AZ 85721, USA\\
\llap{$^c$}
        Physics Department, Washington University\\
St. Louis, MO 63130, USA\\
\llap{$^d$}
        Physics Department, University of Utah\\
Salt Lake City, UT 84112, USA\\
\llap{$^e$}
        American Physical Society\\
 One Research Road, Box 9000, Ridge, NY 11961-9000, USA\\
\llap{$^f$}
        Physics Department, University of the Pacific\\
 Stockton, CA 95211, USA\\
\llap{$^g$}
	Argonne Leadership Computing Facility, Argonne National Laboratory\\
	Argonne, IL 60439, USA\\
\llap{$^h$}
        Physics Department, University of California\\
 Santa Barbara, CA 93106, USA\\
        E-mail: \email{sg AT indiana.edu}
}

\abstract{
We present our new results for the QCD equation of state at nonzero
chemical potential at $N_t=6$ and compare them with 
$N_t=4$.  We use the Taylor expansion method
with terms up to sixth order in simulations with 2+1 flavors of improved
asqtad quarks along a line of constant physics with $m_l=0.1 m_s$ and
approximately physical strange quark mass $m_s$.}

\FullConference{The XXVI International Symposium on Lattice Field Theory \\
                 July 14 - 19, 2008\\
                 Williamsburg, Virginia, USA}

\begin{document}

\section{Introduction}
Experiments at RHIC start with a baryon rich environment; hence they
naturally have a non-zero chemical potential.
The finite temperature field theory formalism easily admits a chemical
potential; however, it leaves us with a complex action, and we can no longer
use importance sampling.  This results in the well known sign problem.
If the chemical potential is small, we can employ the Taylor expansion method
\cite{taylor}.  This method requires simulations only at zero chemical
potential.  Here we employ this method to study QCD with three dynamical
quarks using the asqtad action \cite{Asqtad} that we have already
extensively studied at non-zero temperature, but without a chemical
potential.  Previously, we have performed a study with nonzero
chemical potential only for $N_t=4$ \cite{nt4chem}.
%Similar calculations have been presented at this conference by 
%Ph.~de~Forcrand, S.~Gupta, K.~Kanaya, A.~Nakamura, and C.~Schmidt.

\section{Methodology}
\begin{figure}
\begin{center}
\begin{tabular}{c c}
\includegraphics[width=0.45\textwidth]{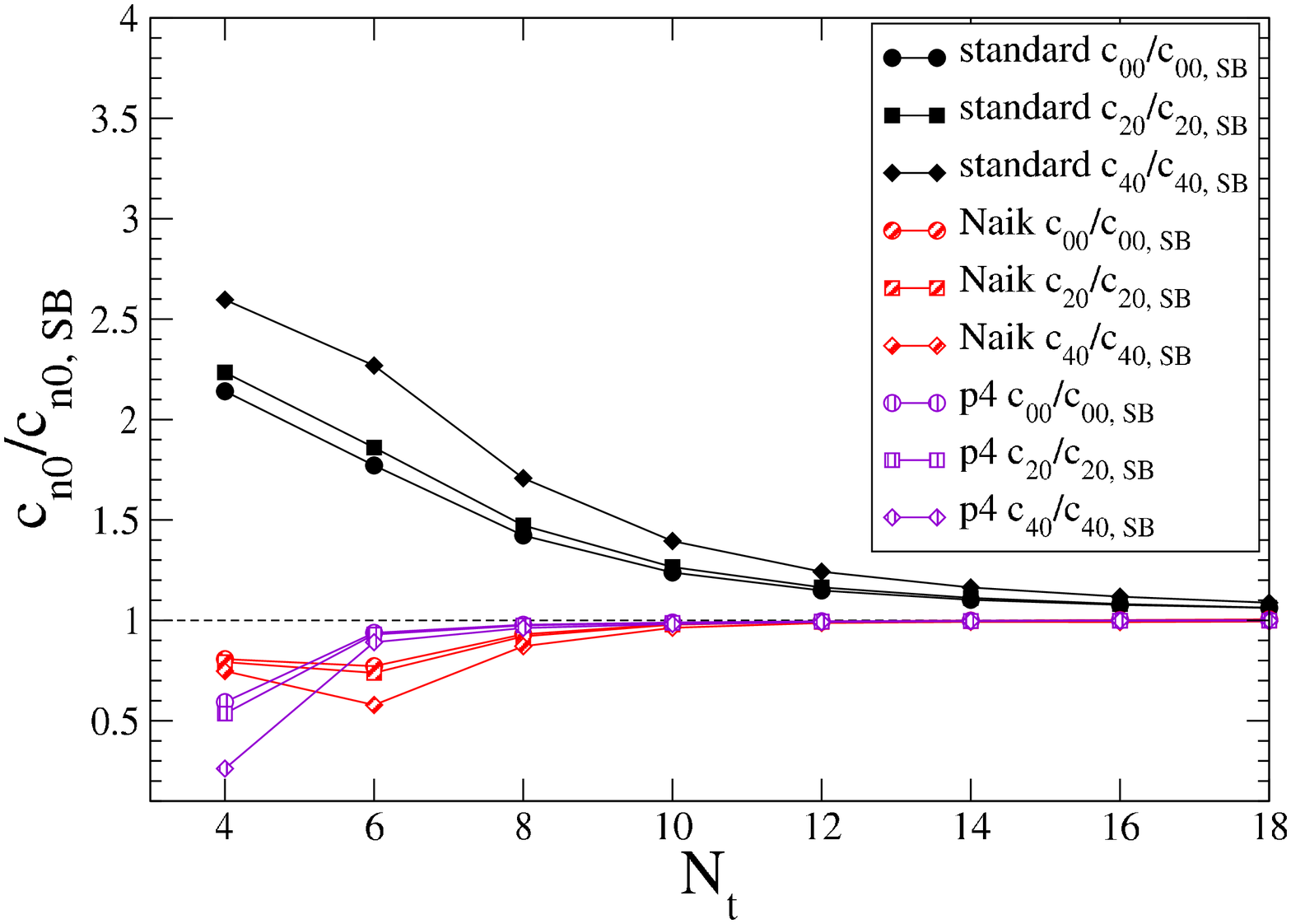}
&
\includegraphics[width=0.45\textwidth]{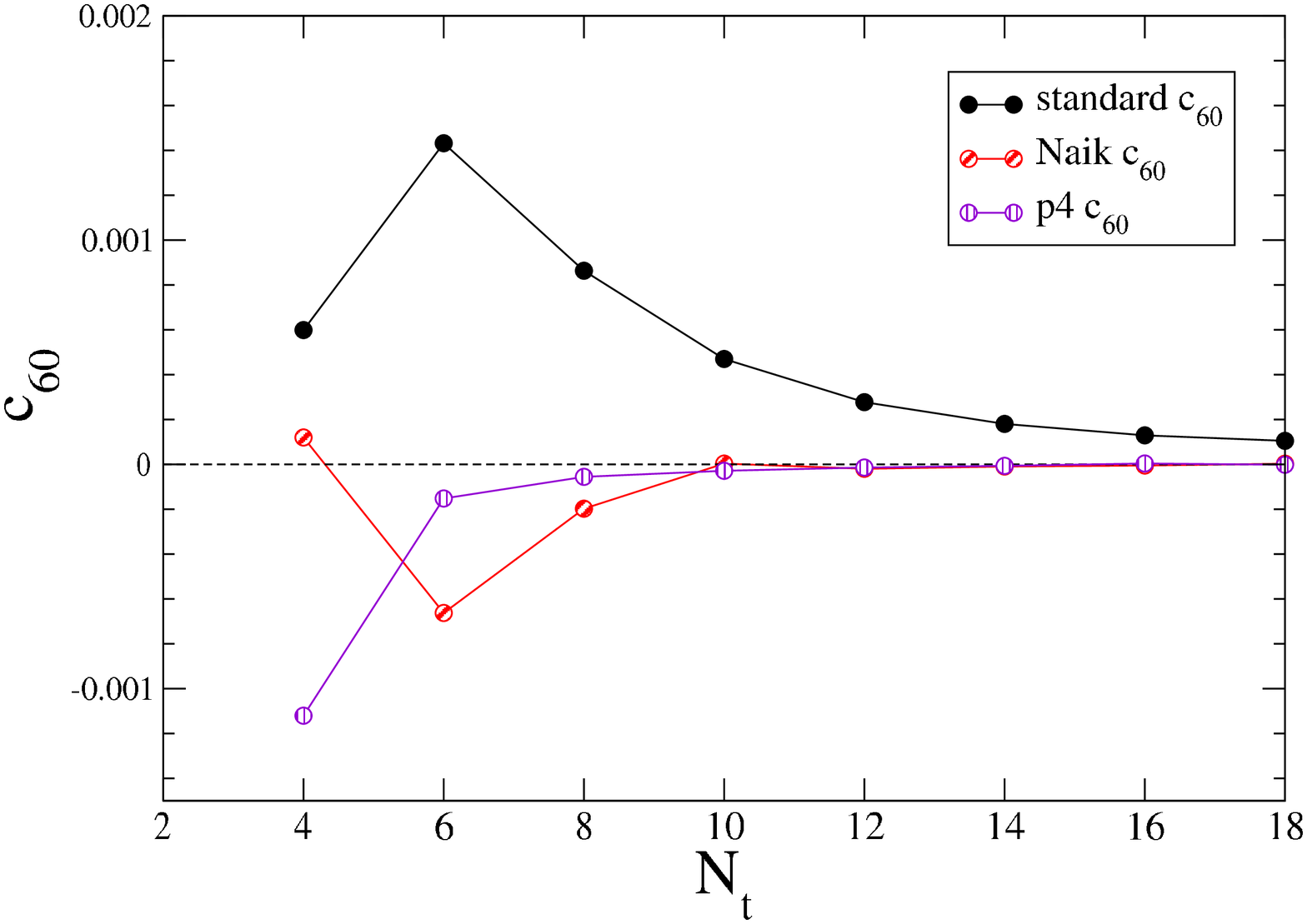}
\end{tabular}
\end{center}
\caption{Several expansion coefficients as a function of $N_t$ for the
free theory \cite{nt4chem}.  On the
left, for non-vanishing coefficients, we show the the ratio
to the Stefan-Boltzmann limit.  On the right, for $c_{60}$, we show the
value itself.}
\label{fig:free}
\end{figure}

We briefly review the formalism and methods that are detailed in
Refs.~\cite{taylor} and \cite{nt4chem}.
Physical quantities of interest are Taylor expanded in the chemical potentials
(in physical units)
$\bar{\mu}_l$ and $\bar{\mu}_h$ for light and strange quarks,
respectively.  We drop the bar when referring to the chemical potential in
lattice units.
For example:
\BE
{p\over T^4}={\ln Z \over VT^3}=
\sum_{n,m=0}^\infty c_{nm}(T) \left({\bar{\mu}_l\over T}\right)^n
\left({\bar{\mu}_h\over T}\right)^m.\label{eq:p}
\EE
Only terms with $n+m$ even appear due to $CP$ symmetry.
\BE
c_{nm} (T)=
{1\over n!}{1\over m!}{N_\tau^{3}\over N_\sigma^3}{{\partial^{n+m}\ln{Z}}\over
{\partial(\mu_l N_\tau)^n}{\partial(\mu_h N_\tau)^m}}\biggr\vert_{\mu_{l,h}=0} \quad.
\label{eq:cn}
\EE
For the interaction measure,
\BE
{I\over T^4}=-{N_t^3\over N_s^3}{d\ln Z \over d\ln a}=\sum_{n,m}^\infty
b_{nm}(T)\left({\bar{\mu_l}\over T}\right)^n
\left({\bar{\mu_h}\over T}\right)^m.
\EE
Temperature dependendent coefficients $c_{nm} (T)$ and $b_{nm} (T)$
are combinations of observables that can be calculated on non-zero
$T$ ensembles, but with zero chemical potential.
We Taylor expand up to sixth order.  To compute all the required terms,
40 fermionic observables have to be determined using stochastic estimators, as
well as several gluonic observables \cite{nt4chem}.
Ensembles are generated on a line of constant physics with $m_l=0.1 m_s$
and $m_s$ approximately the physical strange quark mass.
Our previous work used lattices with $N_t=4$.  We now use $N_t=6$
and compare with the coarser lattices.  Before we present our results,
it is interesting to compare the free theory for different $N_t$ to
see how the continuum limit is approached.  
(See Fig.~\ref{fig:free} \cite{nt4chem}.)

\begin{figure}
\begin{center}
\includegraphics[width=0.95\textwidth]{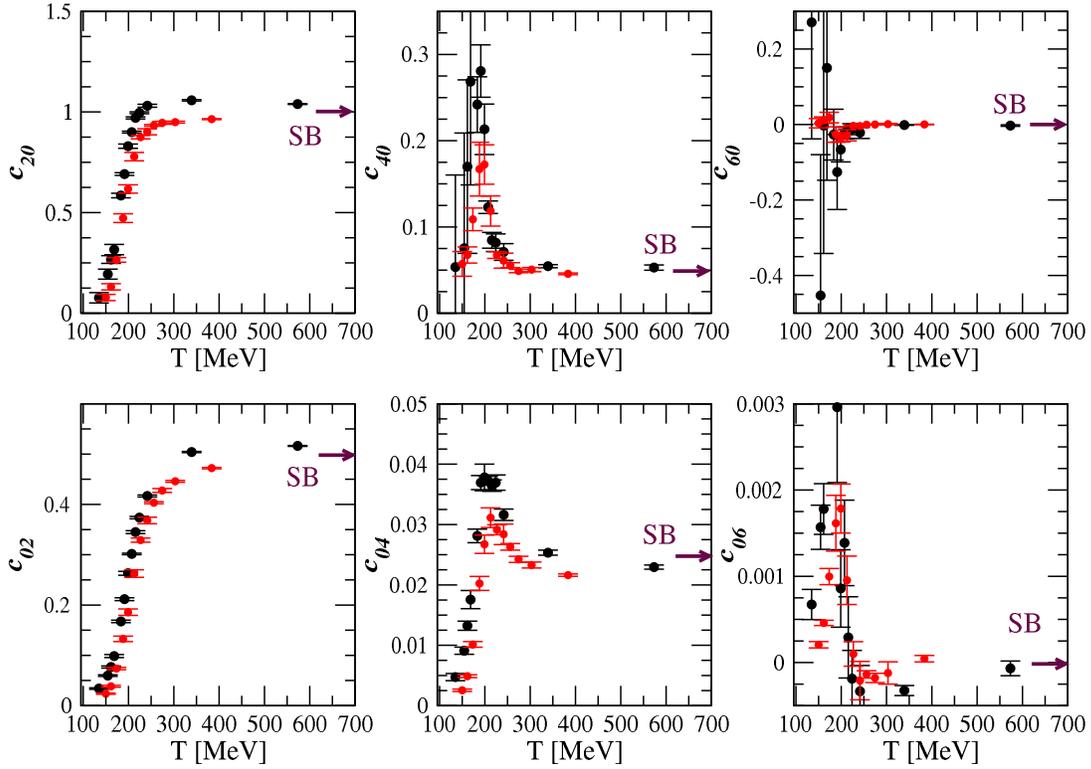}
\end{center}
\caption{Unmixed Taylor coefficients $c_{n0}$ and $c_{0n}$ as a function
of temperture.
New results for $N_t=6$ are shown in red; black is used for $N_t=4$.
\label{fig:unmixed}}
\end{figure}

Turning to the interacting theory we show the unmixed coefficients 
for the pressure in Fig.~\ref{fig:unmixed}.
There is considerable structure at low $T$ and then an approach to the
Stefan-Boltzmann (SB)
limit above the cross-over temperature. Also, the higher order
coefficients are small, but their errors grow rapidly.  Note that the
errors are better controlled for $N_t=6$ (red) than they were 
for $N_t=4$ (black).

\begin{figure}
\begin{center}
\includegraphics[width=0.98\textwidth]{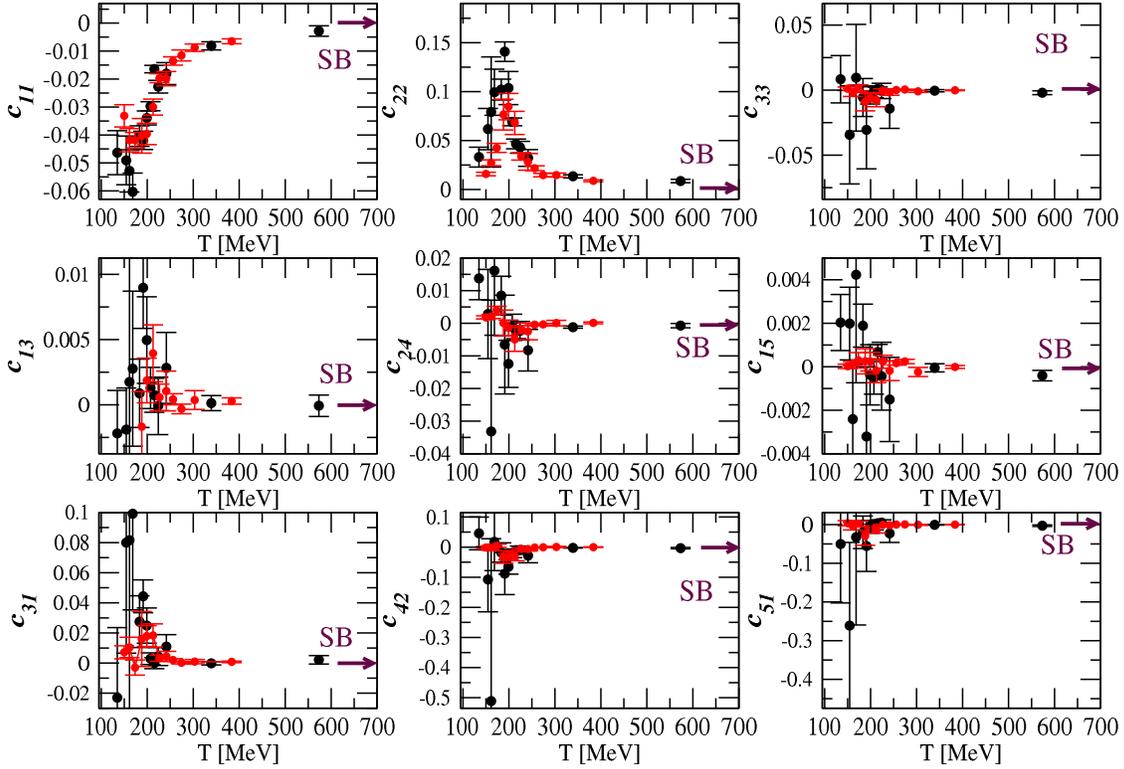}
\end{center}
\caption{Mixed Taylor coefficients $c_{mn}$ as a function
of temperture.
New results for $N_t=6$ are shown in red, black is used for $N_t=4$.}
\label{fig:mixed}
\end{figure}

In Fig.~\ref{fig:mixed}, we show the mixed coefficients for the pressure.
Similar figures are available for the coefficients that are relevant
for the interaction measure.  Due to lack of space, we will not show them
here.

\section{Results}
With the coefficients in hand, we can calculate interesting quantities, such
as pressure, interaction measure, energy density, quark number density, and
quark number susceptibility.
Due to non-zero $c_{n1}(T)$ terms a non-zero strange quark density is induced
even with $\mu_h=0$.  To study the $n_s=0$ plasma, we must, therefore,
tune  $\mu_h$ as a function of $\mu_l$ and $T$.

\begin{figure}
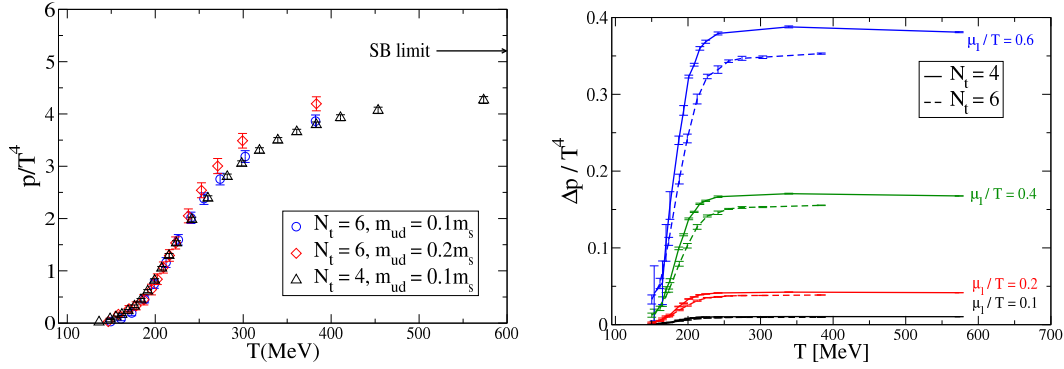

\begin{center}
\begin{tabular}{c c}
\includegraphics[width=0.45\textwidth]{P_pbp_corr.eps}
&
\includegraphics[width=0.45\textwidth]{dP.eps}
\end{tabular}
\end{center}
\caption{The pressure with zero chemical potential (left) and 
the change in pressure due to chemical potential (right).}
\label{fig:dp}
\end{figure}

\begin{figure}
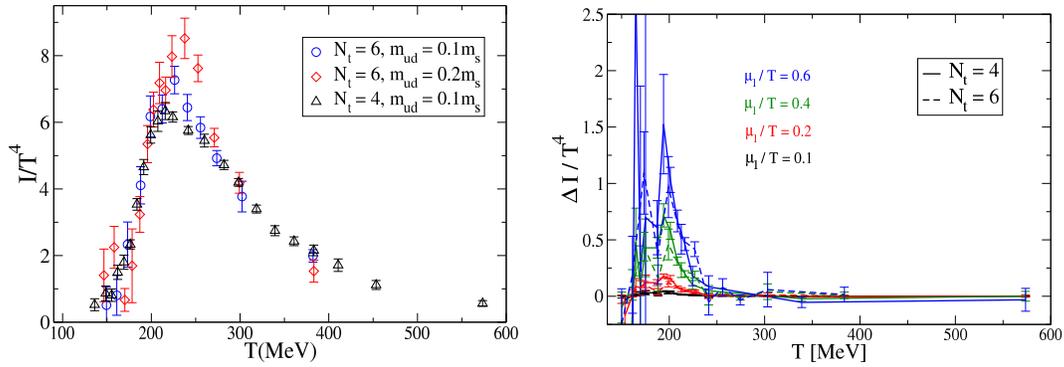

\begin{center}
\begin{tabular}{c c}
\includegraphics[width=0.45\textwidth]{I_pbp_corr.eps}
&
\includegraphics[width=0.45\textwidth]{dI.eps}
\end{tabular}
\end{center}
\caption{The interaction measure with zero chemical potential (left) and 
the change in interaction measure due to chemical potential (right).}
\label{fig:di}
\end{figure}

\begin{figure}
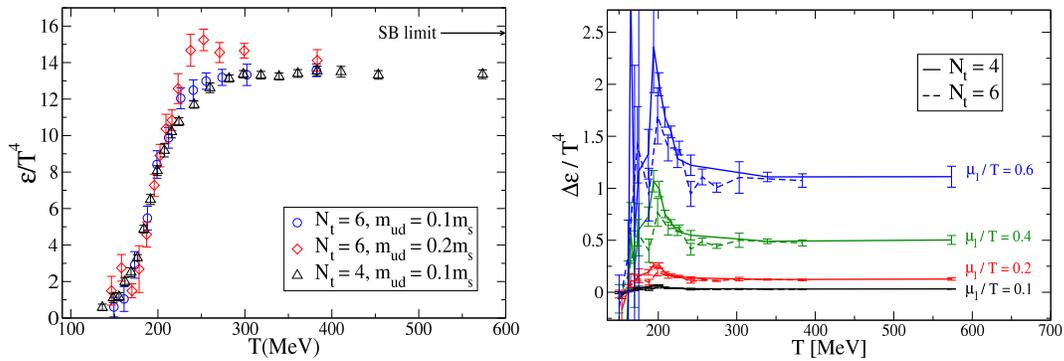

\begin{center}
\begin{tabular}{c c}
\includegraphics[width=0.45\textwidth]{E_pbp_corr.eps}
&
\includegraphics[width=0.45\textwidth]{dE.eps}
\end{tabular}
\end{center}
\caption{The energy density with zero chemical potential (left) and 
the change in energy density due to chemical potential (right).}
\label{fig:de}
\end{figure}

\begin{figure}
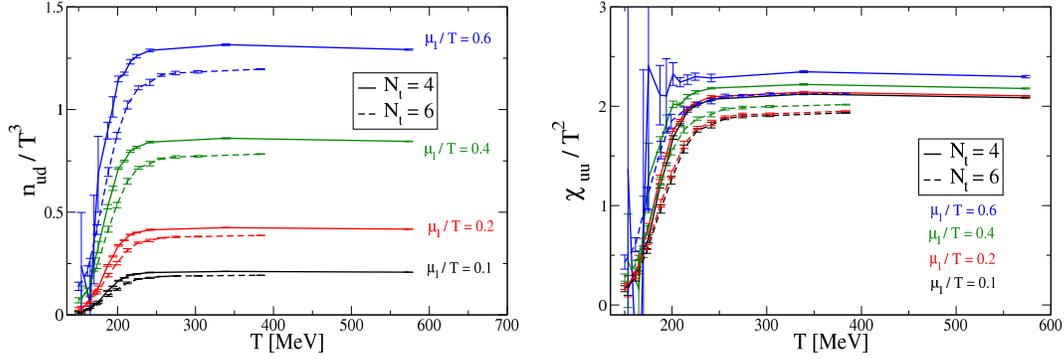

\begin{center}
\begin{tabular}{c c}
\includegraphics[width=0.45\textwidth]{dNud.eps}
&
\includegraphics[width=0.45\textwidth]{dXuu.eps}
\end{tabular}
\end{center}
\caption{Light quark number density (left) and quark number 
susceptibility (right).}
\label{fig:numberdensity}
\end{figure}

In Figs.~\ref{fig:dp} and \ref{fig:di}, we show how the pressure 
and interaction measure change as a function of $T$ for selected
values of $\mu/T$.  In each figure, we display the zero chemical potential
case on the left.  We compare results for $N_t=6$ with our
prior results for $N_t=4$.  We find that the change in pressure is somewhat
smaller compared with our previous results.  For the interaction measure,
the errors are fairly large, but there also seems to be a reduction there.
Figure~\ref{fig:de} shows the energy density and change in energy density
due to chemical potential.  In this case, the differences between $N_t=4$ and
6 are small.
In Fig.~\ref{fig:numberdensity}, we show the light quark number density and
the quark number susceptibility.  We note that both the number density and
susceptibility are somewhat smaller with $N_t=6$ than they were for $N_t=4$.

\begin{figure}
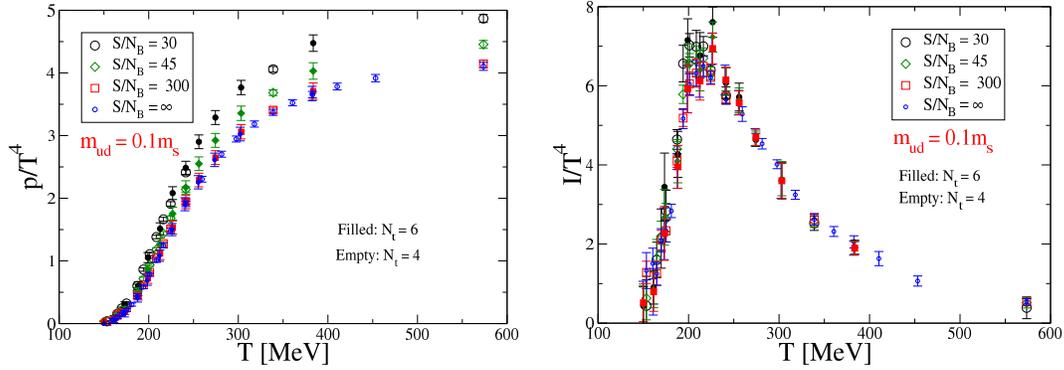

\begin{center}
\begin{tabular}{c c}
\includegraphics[width=0.45\textwidth]{PO6isentr.eps}
&
\includegraphics[width=0.45\textwidth]{IO6Isentr.eps}
\end{tabular}
\end{center}
\caption{Isentropic pressure (left) and interaction measure (right) for
selected values of $S/n_B$ appropriate to AGS, SPS and RHIC.}
\label{fig:isentropic}
\end{figure}

\begin{figure}
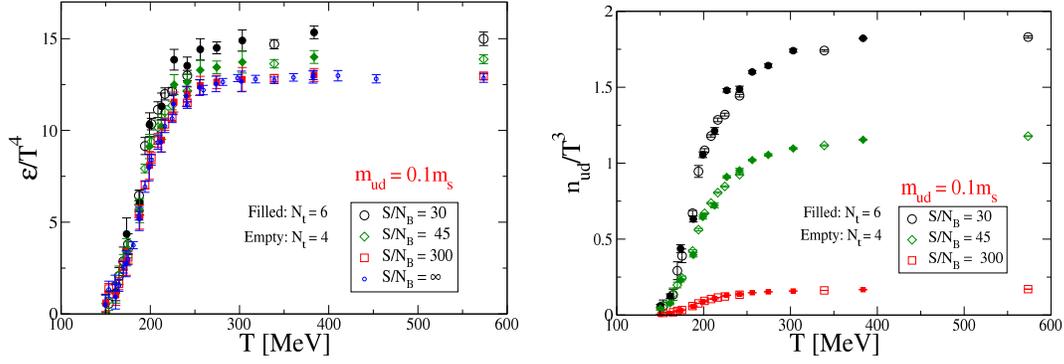

\begin{center}
\begin{tabular}{c c}
\includegraphics[width=0.45\textwidth]{EO6Isentr.eps}
&
\includegraphics[width=0.45\textwidth]{nuO6Isentr.eps}
\end{tabular}
\end{center}
\caption{Isentropic energy density (left) and light quark number 
density (right) for
selected values of $S/n_B$ appropriate to AGS, SPS and RHIC.}
\label{fig:isentropic2}
\end{figure}

\begin{figure}
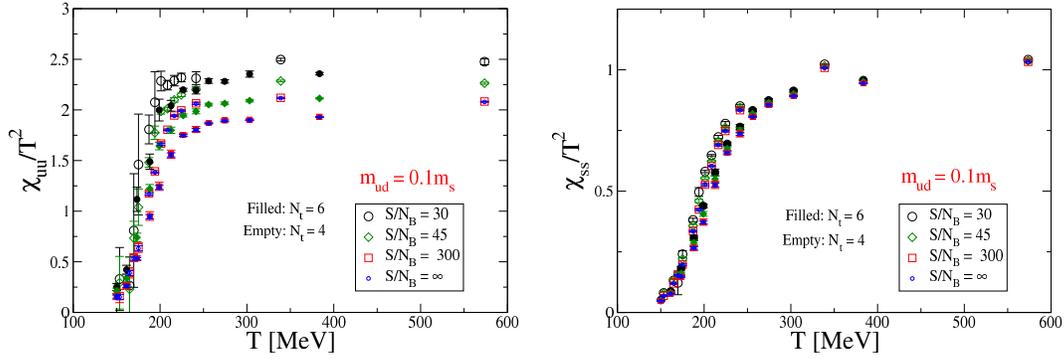

\begin{center}
\begin{tabular}{c c}
\includegraphics[width=0.45\textwidth]{Chi_uuO6Isentr.eps}
&
\includegraphics[width=0.45\textwidth]{Chi_ssO6Isentr.eps}
\end{tabular}
\end{center}
\caption{Isentropic light quark number susceptibility (left) and strange 
quark number susceptibility (right) for
selected values of $S/n_B$ appropriate to AGS, SPS and RHIC.}
\label{fig:isentropic3}
\end{figure}

One particularly interesting quantity is the isentropic equation of state (EOS).
In a heavy-ion collision, after thermalization, the system expands and
cools with constant entropy.  Thus,
we would like to find the EOS with fixed ratio of entropy to baryon number.
The appropriate ratio of $s/n_B$ for AGS, SPS and RHIC
are 30, 45 and 300, respectively.
To carry out this calculation we must find 
trajectories in the ($\mu_l$,$\mu_h$,$T$) space
with $n_s=0$ and $s/n_B$ as stated above.  In Fig.~\ref{fig:isentropic}, 
we show the isentropic pressure and interaction measure.
We also show the isentropic energy density, and quark number density in
Fig.~\ref{fig:isentropic2}, and the isentropic
light and strange quark susceptibilities in Fig.~\ref{fig:isentropic3}.

\section{Conclusions}
We have extended our sixth order Taylor expansion
study of thermodynamics with chemical potential toward
the continuum limit by going from $N_t=4$ to $6$.
After computing the expansion
coefficients relevant for both pressure $p$ and
interaction measure $I$ we can compute a number of interesting quantities.
We observe modest lattice spacing effects, 
with the quark densities and susceptibilities, and the 
effect of chemical potential, smaller 
at the smaller lattice spacing.
In addition, 
we  have calculated the isentropic equation of state, which is particularly
relevant for the phenomenology of relativistic heavy-ion colliders.
For both values of $N_t$ we find rather smooth behavior for the isentropic
variables indicating that experiments are far from any critical point
in the $\mu$--$T$ plane.

It would be interesting to extend this work to yet smaller lattice spacing
and to go to lighter quark mass.

\section*{Acknowledgements}
This work was supported by the U.S. Department of Energy and the National
Science Foundation.

\end{document}